\documentclass[lettersize,journal]{IEEEtran}
\usepackage{amsmath,amsfonts}
\usepackage{algorithmic}
\usepackage{algorithm}
\usepackage{array}
\usepackage[caption=false,font=normalsize,labelfont=sf,textfont=sf]{subfig}
\usepackage{textcomp}
\usepackage{multirow}
\usepackage{stfloats}
\usepackage{verbatim}
\usepackage{graphicx}
\usepackage{import}
\usepackage{circuitikz}
\usepackage{makecell}
\usepackage[backend=biber,style=ieee]{biblatex}
\usepackage{hyperref}
\usepackage[hyphenbreaks]{breakurl}
\usepackage{xurl}
\hypersetup{
    colorlinks=true,
    linkcolor=black,
    filecolor=black,      
    urlcolor=blue,
    citecolor=black,
    pdftitle={A Linear Pricing Mechanism for Load Management in Day-Ahead Retail Energy Markets}
}

\begin{document}

% \title{Load-Responsive Pricing for Distribution System Congestion Management in Day-Ahead Retail Energy Markets}
\title{A Linear Pricing Mechanism for Load Management in Day-Ahead Retail Energy Markets}

\author{Phillippe K. Phanivong,~\IEEEmembership{Member,~IEEE}, Duncan S. Callaway,~\IEEEmembership{Member,~IEEE}
        % <-this % stops a space
\thanks{P. K. Phanivong and D. S. Callaway are with the Energy and Resources Group,
University of California, Berkeley, CA 94720-3050 USA (e-mail: phillippe\_phanivong@berkeley.edu; dcal@berkeley.edu).}}

\maketitle

\begin{abstract}
Regulators and utilities have been exploring hourly retail electricity pricing, with several existing programs providing day-ahead hourly pricing schedules.  At the same time, customers are deploying distributed energy resources and smart energy management systems that have significant flexibility and can optimally follow price signals. In aggregate, these optimally controlled loads can create congestion management issues for distribution system operators (DSOs). In this paper, we describe a new linear pricing mechanism for day-ahead retail electricity pricing that provides a signal for customers to follow to mitigate over-consumption while still consuming energy at hours that are preferential for system performance. We show that by broadcasting a linear price designed for price-signal control of cost-optimizing loads, we can shape customer load profiles to provide congestion management without the need for bi-directional communication or customer bidding programs. 
% We show that by creating a small linear pricing mechanism designed for price-signal control of cost-optimizing loads, we can shape customer load profiles to provide congestion management with minor increases in customer bills that can later be returned for revenue neutrality or used in a subscription mechanism.
\end{abstract}


\vspace{-5mm}
\section{Introduction}
\label{sec:intro}
\IEEEPARstart{A}{s} more flexible loads and distributed energy resources (DERs) are being adopted by customers, there is a growing interest in retail tariffs that provide real-time pricing \cite{CalFuse_Report_2022}, \cite{FERC_DR_Assessment_2022}. In practice, most of these tariffs are day-ahead hourly pricing schemes \cite{EPRI_RTP_2021}. This type of pricing mechanism incentivizes customers to shift load to better follow prices on the wholesale market, reduce peak load on the system, and reduce costs. However, these pricing systems often do not take into account the effects on the distribution system. 

% At the same time more DERs are coming onto the distribution system, communication and load control technologies are improves leading to customers better controlling and optimizing their loads [INSERT A CITIATION OR TWO ON LOAD CONTROL TECH]. As a result, a new class of load is beginning to develop that is similar to the thermostatically controlled load but is based on optimally consuming to minimize costs for the customer. These optimally controlled loads (OCLs) include building energy management systems, smart thermostats that are price-aware, and electric vehicles that charge or potentially discharge when prices are favorable for the customer. 

As more customers install smart energy management systems and price-responsive loads, system operators may start to see new potential issues for the power grid, especially at the local distribution level. Sufficient distribution system capacity is already a concern for electric vehicle adoption and other distributed energy resources (DERs) today \cite{Elmallah_Brockway_Callaway_2022}. Under day-ahead prices, customers are given a schedule of time-varying volumetric energy prices for the following day. These volumetric prices are usually set at the hourly level, however some markets may change sub-hourly. The price schedule allows customers to optimize their load profile to minimize costs. If enough loads optimize towards the same day-ahead prices, in aggregate they may create new system peaks causing congestion. This congestion can lead to an increase in losses, system undervoltage, or overloading of equipment.

Essentially, regulators and utilities are exploring day-ahead energy pricing so that customers can respond to prices. However, when enough customers shifting load to optimize for their prices causes grid congestion, prices will need to respond to load. At the wholesale level, this is done by using real-time markets in addition to the day-ahead market. Some researchers have proposed similar markets for the distribution system \cite{Ding_Pineda_Nyeng_Østergaard_Larsen_Wu_2013}. However, real-time market prices can be more challenging to optimize against because there is less time for customers to solve their optimization problems, and there is more uncertainty about prices in future hours. In addition, market operators need real-time visibility into grid conditions and fast algorithms to compute prices are spatially differentiated to address distribution system congestion. A 2021 ERPI analysis of the 55 available real-time and day-ahead retail pricing programs in the US found that only four of the programs use spatially differentiated real-time or day-ahead prices \cite{EPRI_RTP_2021}. 

While most utility companies in the U.S. do not use day-head or real-time pricing, they do use other pricing mechanisms to reduce customer peak load that could be integrated with day-head or real-time pricing. Some utility tariffs use demand charges (monthly costs based on the peak power consumption of a customer) to recover capacity costs from customers \cite{NREL_Demand_Charges_2017}. These charges can cause customers to reduce their peak load \cite{Powell2020Controlled}. However, demand charges were not designed for congestion management and are criticized for their economic impact, since individual customer peaks may not correlate with times of system peaks \cite{Borenstein2016economics}.  In addition, at high adoption levels of price-optimizing loads, customers are incentivized by demand charges to flatten their daily load profile instead of responding to time-varying energy prices \cite{Phanivong_EV_Rates_2023}.  

Another way utilities can reduce peak load is through Critical Peak Pricing (CPP) programs. These programs alert customers to conserve energy on specific days, called CPP days. However, CPP programs are typically designed to address transmission level congestion and are limited in the number of CPP days they can call per year \cite{Schittekatte_et_al_2022}.

Researchers have examined several ways to perform congestion management outside of demand charges and CPP, both through direct load control and market based approaches. In the context of day-ahead energy pricing, direct load control may interfere with customer decisions and those of third party aggregators. However, market methods for DER congestion management focus on changing customer behavior and decision making based on costs and price signals. 

A rich area of research has been transactive energy (TE) markets \cite{hao_transactive_2017}\nocite{li_distributed_2018}-\cite{Ullah_Park_2023}. TE markets provide prices to customers and producers to trade energy as needed to maximize net benefits. The authors of \cite{hao_transactive_2017} use transactive control to provide demand response with commercial building loads. While in \cite{li_distributed_2018}, the authors extend a transactive framework to provide distributed control of distribution networks. State of the art TE markets work includes \cite{Ullah_Park_2023}, where the authors develop a decentralized energy market for coordinating between transmission system operators, distribution system operators (DSOs), and the individual distributed energy resource (DER) owner.

Similar to the TE markets, researchers have explored various bidding-based control schemes that require iterative, bi-directional communication between the DSO and customers. In \cite{Xu_Li_Low_2016}, customers submit supply functions to the DSO and the DSO establishes a clearing price. 

An equally mature research area has been the development of Distribution Locational Marginal Pricing (DLMP) approaches to congestion management \cite{sotkiewicz_nodal_2006}\nocite{li_distribution_2014}\nocite{caramanis_co-optimization_2016}\nocite{mohammadi_distribution_2022}-\cite{zhao_distribution_2023}. While nodal pricing in the distribution system has been discussed for many years, the authors in \cite{sotkiewicz_nodal_2006} developed one of the first formulations. DLMP has since been examined for EV charging management \cite{li_distribution_2014},  co-optimization of power and reserves \cite{caramanis_co-optimization_2016}, and management of markets under uncertainty \cite{zhao_distribution_2023}.

Other market-based mechanism that are actively being researched include incentive mechanisms such as dynamic subsidies paid by the DSO to customers \cite{huang_dynamic_2018}. Similar incentives are proposed as a Stackelberg game for bi-level optimization of the DSO and customers in \cite{fattaheian-dehkordi_incentive-based_2021} and as a Nash-Stackelberg-Nash game for EV coordination in \cite{lv_coordinating_2023}.

Although researchers have explored many different price-signal calculation mechanisms, each of the price signals in the papers referenced above lead to a total customer cost that is linear in quantity. This type of problem can be solved by each customer via a linear program. However, in \cite{huang_distribution_2015}, Huang et. al. explore quadratic energy costs and identify quadratic programming as a superior approach to linear programming due to the multiple solutions a customer or aggregator may arrive at. They extend this work to provide a dynamic power tariff (DPT) in \cite{huang_dynamic_2019}. In both papers, the DSO is required to calculate quadratic energy costs based on system conditions, either as a DLMP \cite{huang_distribution_2015} or as part of the DPT \cite{huang_dynamic_2019}. However, calculating a spatially granular volumetric energy price can be time consuming for a DSO or might not be possible if the DSO does not have measurements of system conditions. 

To address this problem, we separate the calculation of quadratic costs into a volumetric energy price, similar to conventional day-head prices, and a pricing component that is linearly dependent on quantity. This composite price, referred to in this paper as Load-Responsive Pricing (LRP), is then communicated to customers as the slope and intercept of a linear pricing curve for each pricing period in the day-ahead schedule. This creates a quadratic cost curve for the customer to optimize against in the day-ahead market, building in a cost for congestion management similar to the quadratic pricing used in \cite{huang_distribution_2015}. What separates our LRP approach from earlier work is that our quadratic costs are supplemental to traditional volumetric energy prices, and we do not need to take into account real-time system conditions to construct these prices. This approach allows DSOs to determine day-ahead energy prices and optimal load profiles of price-responsive loads independently. Then, once both have been determined, the DSO can use LRP as a price-signal to drive customers to the desired load shape as a target load profile under any volumetric energy price. In addition, for situations in which DSOs have insufficient information to construct distribution-system-optimal load profiles, we provide a heuristic method to utilize LRP for congestion management without communication between the customer and the DSO beyond transmitting meter data and prices. It should be noted that LRP, as with all price-signal control approaches, increase costs for the customer compared to direct control. As such, we describe several ways to minimize these cost increases and to reimburse the customer for this control scheme if necessary.

\subsubsection*{\bf The three main contributions of this work are}
\begin{enumerate}
\item{We propose a quadratic cost mechanism, Load-Responsive Pricing (LRP), for price-signal control of flexible loads and we demonstrate its performance for distributed congestion management in a large, three-phase unbalanced distribution feeder.}
\item{We derive an optimal pricing strategy for LRP and show how it can shape customer load profiles to match target load profiles under any day-ahead pricing schedule.}
\item{We also develop a heuristic pricing algorithm for congestion management using LRP without the need for optimal powerflow analysis.}
\end{enumerate}

The organization of the rest of this paper is as follows: Section \ref{sec:LRP} describes the linear pricing mechanism we propose. Next, in Section \ref{sec:Optimal-LRP}, we derive a formula for LRP price-setting based on load profiles generated by the DSO. Then we present a heuristic pricing algorithm in Section \ref{sec:IR-LRP}. Next, we examine the effectiveness of the LRP mechanism through case studies in Section \ref{sec:case-studies}. We then discuss considerations a DSO or regulator would need to consider to implement a LRP tariff in Section \ref{sec:considerations} and conclude the paper in Section \ref{sec:conclusion}.

\section{Load Responsive Pricing}
\label{sec:LRP}
\subsection{Concept}
We assume a DSO provides hourly or sub-hourly volumetric prices for energy in a 24-hour day-ahead schedule, shown in \eqref{eq:realtime_cost_eq}. In this equation, $x_t$ is the load in kWh, $\beta_t$ is the volumetric price in \$/kWh, and $C_t$ is the volumetric cost of energy in dollars at time $t$, and where $t$ is the interval used for the day-ahead price. The most common time interval duration is hourly. Less common time intervals include 5-min and 15-min. We make no assumptions on the source of the $\beta_t$ prices; these prices can be calculated through DLMPs, scaled wholesale prices, or any other means. Customers may also be exposed to other fixed charges in this tariff. Since fix charges would not impact the response of price-signal controlled loads, we exclude these costs from the analysis in this paper to focus on volumetric costs.  

\begin{equation}
\label{eq:realtime_cost_eq}
C_t = x_t  \cdot \beta_t 
\end{equation}

% Assume a customer has a day-ahead schedule of hourly $\beta_t$ prices for $t\in T = 1 \rightarrow 24$. If there are no constraints on when or how much energy that customer can consume, the optimization problem that customer needs to solve is equivalent to determining the minimum $\beta_t$. 

% \begin{equation}
% \label{eq:unconstrained_min}
% \min_{x_t} C_T = \sum_{t=1}^{24} x_t \cdot \beta_t, \quad t\in T= 1\rightarrow24
% \end{equation}

\begin{figure}[!htbp]
\centering
\subfloat[]{\includegraphics[width=0.50\columnwidth]{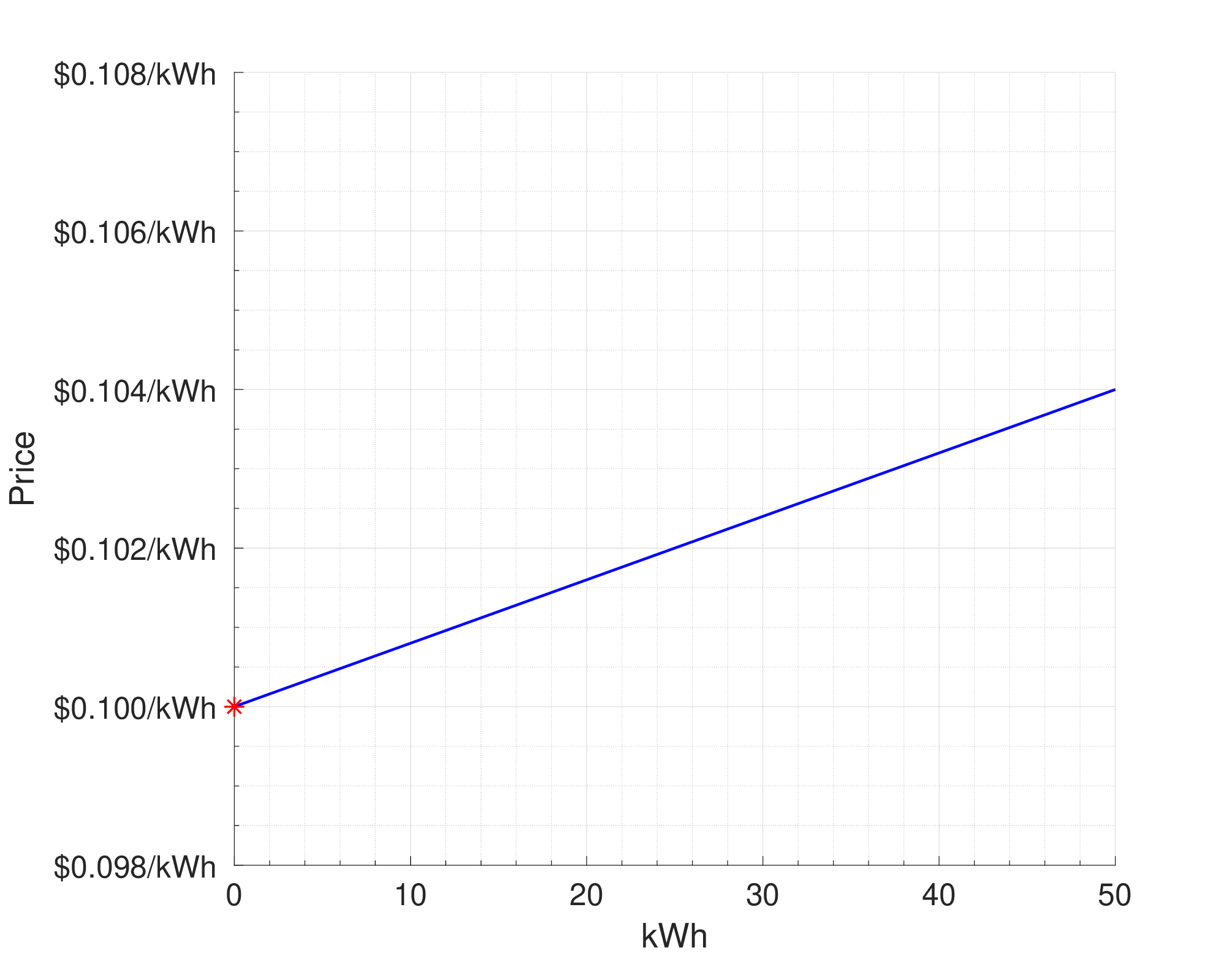}%
\label{fig_1a}}
\hfil
\subfloat[]{\includegraphics[width=0.50\columnwidth]{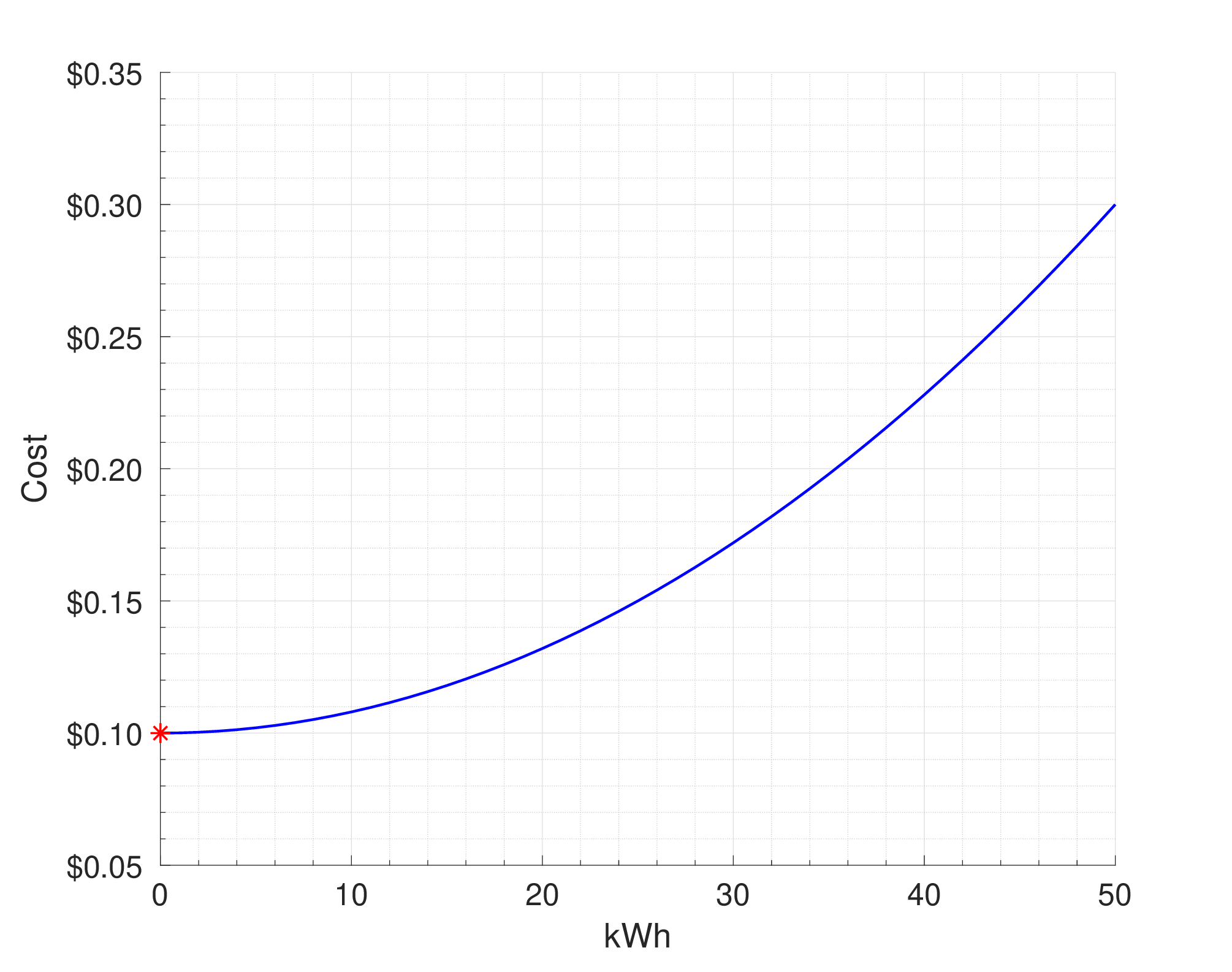}%
\label{fig_1b}}
\caption{A LRP curve for energy at time $t$ with $\beta_t$ intercept and $\alpha_t$ slope. The value for $\alpha_t$ is exaggerated here for visual clarity. (a) The $\pi_t$ pricing curve. (b) The total volumetric cost $C_t$ for the time period.}
\label{fig:LRP_basis}
\end{figure}

We propose adding a linear pricing component to the dynamic tariff, where the linear component depends on the energy consumed in the given period $t$. This new combined price becomes a linear pricing curve $\pi_t$ with a slope component $\alpha_t$ that controls the rate of price increases based on consumption, and an intercept $\beta_t$ that is the time-varying price of energy \eqref{eq:LRP_price_eq}. The purpose of this $\alpha_t$ component is to place a price on congestion in a given period. Equation \eqref{eq:LRP_cost_eq} is the total volumetric LRP cost of energy consumption at time $t$. Fig. \ref{fig:LRP_basis} shows $\pi_t$ pricing curve and $C_t$ cost curve for a single time $t$.

\begin{equation}
\label{eq:LRP_price_eq}
\pi_t = \alpha_t \cdot x_t + \beta_t
\end{equation}

\begin{equation}
\label{eq:LRP_cost_eq}
C_t = x_t  \cdot \pi_t = x_t \left(\alpha_t \cdot x_t + \beta_t\right) = \alpha_t x_t^2 +\beta_t x_t
\end{equation}

\vspace{-2.5mm}
\subsection{Setting \texorpdfstring{$\alpha$}{α} in LRP}
\begin{figure}[!htbp]
\centering
\includegraphics[width=3.25in, height = 2.25in]{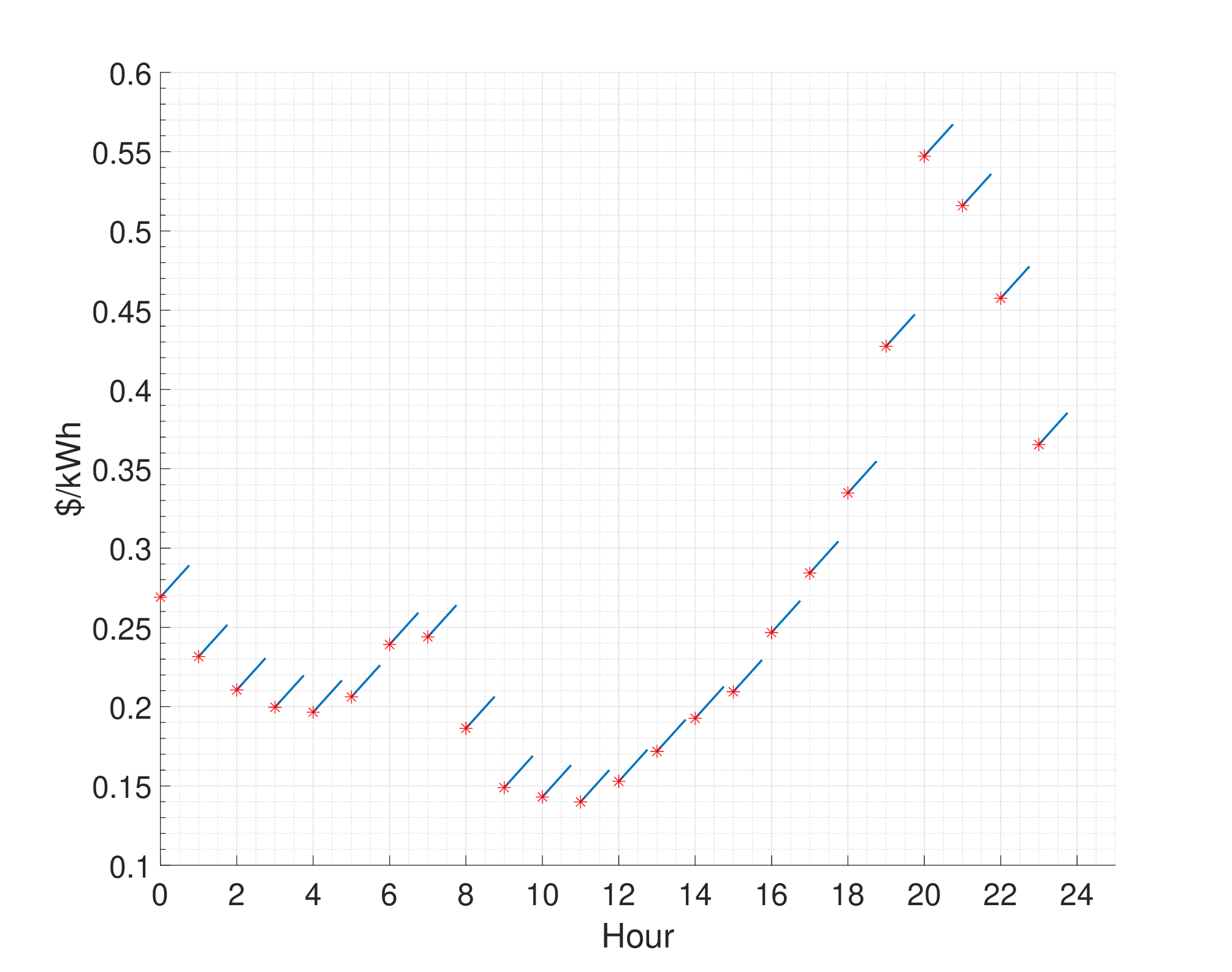}
\caption{A Constant-$\alpha$, hourly LRP. The intercept at each hour is the $\beta_t$ (red asterisk) while the price is calculated with a slope of $\alpha$ for all times $t$.}
\label{fig:Constant_alpha}
\end{figure}

In this paper we assume that customers will adjust the timing of their controllable load to minimize their total cost $C_t$ and that $\beta_t$ prices are greater than zero. Under these assumptions, by choosing $\alpha_t$, a DSO can manage how much a customer will consume at time $t$. In a system without grid congestion, $\alpha_t = 0$ and the customer would just experience the volumetric energy price $\beta_t$. However, if congestion is a concern, the DSO can limit the consumption at time $t$ of a customer by setting $\alpha_t > 0$. The simplest way to set $\alpha_t$ for LRP is to set a constant $\alpha_t$ for all hours of the day (Fig. \ref{fig:Constant_alpha}). However, using dynamic $\alpha_t$ values for different pricing periods allows the DSO to designate times when congestion is a concern and minimize cost increases at other times. In the following sections we propose two approaches to calculating $\alpha_t$ values based on $\beta_t$ prices and system conditions. The first is an optimal-$\alpha$ approach that can be used if a DSO can determine optimal load profiles for its customers. However, if this is not feasible for the DSO, we also present a heuristic based system pricing in Section \ref{sec:IR-LRP}.

% If the peak power measured in a pricing period is also the same as the energy consumed in the same pricing period, the costs to the customer become quadratic.  

% To create a synchronized demand charge-like LRP, a single hour of the day could be used as the $\alpha$ hour. At all other hours, the alpha would be zero. This would be different from an actual demand charge because the alpha component would be designed to be as small as effective for changing the customer's consumption pattern towards a level that is desirable for grid operations. That level will be dependent on customer size.

% By using $\beta$ for cost recovery, we can minimize the size of $\alpha$ to only the required size for controlling OCL behavior.  

% issue that augments day-ahead dynamic tariffs. Load-Responsive Pricing (LRP) is an approach to create variable energy pricing based on an individual customer's consumption level. The conceptual basis for LRP is relative pricing. For example, if the marginal cost of energy in any hour is more expensive relative to any other hour, an OCL will optimize towards the cheaper hour. As such, even small changes (i.e., fractions of a cent) in this relative pricing can change the load profile of OCLs. 

\section{Optimal-\texorpdfstring{$\alpha$}{α} LRP}
\label{sec:Optimal-LRP}
\subsection{Concept}
If the DSO can accurately forecast each customer's total energy demand over an optimization horizon, they can use the results to calculate $\alpha$ coefficients in the LRP. For example, the DSO use an optimal powerflow analysis to determine the optimal load profile for controllable loads on a distribution feeder. This ``optimal-$\alpha$ LRP" method uses the these load profiles as a target profile for LRP customers and sets $\alpha_t$ values for each customer that will cause the customer to optimize to the target load levels given any set of $\beta_t$ prices.  

The optimal-$\alpha$ LRP shares some similarities to the Dynamic Power Tariff (DPT) proposed in Huang et al. \cite{huang_dynamic_2019}, though the prices are constructed in different ways. Huang et al. build on their quadratic DLMP pricing work \cite{huang_distribution_2015} by proposing a quadratic power tariff with prices determined through an iterative process between the distribution system operator (DSO) and customers (in their paper these customers are DER aggregators), where customers propose load profiles and the DSO updates the DPT prices with marginal network costs calculated from an OPF using dc powerflow. In this approach, the day-ahead prices are calculated concurrently in the same process as the load profiles are negotiated. 

In the optimal-$\alpha$ LRP, we separate the calculation of $\alpha_t$ prices to occur after target load profiles and $\beta_t$ energy prices are calculated. This decoupled approach allows the DSO to use any OPF strategy they wish to identify the desired load profile. At the same time, $\beta_t$ prices can be determined via DLMP, directly using wholesale prices, or some other novel method. Optimal-$\alpha$ LRP is agnostic to the approach to calculating load profiles and $\beta_t$ prices. 

\subsection{Formula}
The optimal-$\alpha$ LRP calculates $\alpha_t$ values for each pricing period $t$ %in the set of $\beta_t\; \forall t \in T$ , 
given a set of target load profiles for customers ($\hat{x}_t$). %Without loss of generality, we assume $t$ to be hourly over a $T = 24$ hour pricing day. 
For clarity, in the following we assume all of the customer load is controllable and separately metered from non-controllable loads. We address issues for customers with both controllable and non-controllable loads on a single meter at the end of this subsection. 

The $\beta_t$ values can represent wholesale market costs, and they could also include additional time-varying costs DSOs wish to pass on to customers, with the considerations described in Section \ref{sec:considerations}. Using the $\beta_t$ prices, each customer is assumed to solve their own cost minimizing quadratic optimization and the $\hat{x}_t$ values calculated by the DSO are assumed to be feasible solutions given each customer's local constraints. Equation \eqref{eq:opt_LRP_eq} represents the quadratic problem customers must solve to minimize their costs, where $x_{1...n}$ is their load at times $t = 1...n$ such that their total load is equal to $X$. 

\begin{align}
\label{eq:opt_LRP_eq}
\min_{x} \;\alpha_1 x_1^2+\beta_1 x_1 &+  \alpha_2 x_2^2 + \beta_2 x_2 +{}...\\ 
... {} &+ \alpha_n x_n^2 + \beta_n x_n \nonumber\\ 
s.t.\; x_1+x_2+...&+x_n = X\nonumber
\end{align}

Since the customer's optimization is quadratic, there is a set of $\alpha_t$ values that produces a unique solution for the customer's optimal load profile given the $\beta_t$ prices, as proven in the similarly formulated quadratic DLMP in \cite{huang_distribution_2015}. By using Lagrange Multipliers, we can calculate these optimal $\alpha_t$ values \eqref{eq:lagrange_eq}-\eqref{eq:lagrange_sol6_eq}.

\begin{align}
\label{eq:lagrange_eq}
\mathcal{L} = \alpha_1 x_1^2 & +\beta_1 x_1 + \alpha_2 x_2^2 + \beta_2x_2 + ... + \alpha_n x_n^2 + \beta_n x_n\\ 
    & + \lambda(X-x_1-x_2-...-x_n) \nonumber     
\end{align}

\begin{equation}
 \label{eq:lagrange_sol1_eq}
\frac{\partial \mathcal{L}}{\partial x_1} = 2\alpha_1 x_1+\beta_1  - \lambda = 0 
\end{equation}

\begin{align}
\label{eq:lagrange_sol2_eq}
\frac{\partial \mathcal{L}}{\partial x_2} &= 2\alpha_2 x_2+\beta_2  - \lambda = 0 \\
&\qquad \vdots \nonumber \\
\frac{\partial \mathcal{L}}{\partial x_n} &= 2\alpha_n x_n+\beta_n  - \lambda = 0 \nonumber
\end{align}

\begin{equation}
 \label{eq:lagrange_sol3_eq}
\frac{\partial \mathcal{L}}{\partial \lambda} = X - x_1 -x_2 - ...- x_n = 0 
\end{equation}

% We rearrange \eqref{eq:lagrange_sol2_eq} and substitute into \eqref{eq:lagrange_sol1_eq} to solve for $\alpha$ \eqref{eq:lagrange_sol4_eq}-\eqref{eq:lagrange_sol6_eq}.

% \begin{equation}
%  \label{eq:lagrange_sol4_eq}
% \lambda = 2\alpha_2 x_2 + \beta_2 
% \end{equation}

% \begin{equation}
%  \label{eq:lagrange_sol5_eq}
% 2\alpha_1 x_1+\beta_1 - 2\alpha_2 x_2 + \beta_2 = 0
% \end{equation}

We can solve for any $\alpha_t$ in terms of its $\beta_t$ and the $\alpha, \beta$, and $x$ values from any other interval.  For example, 
\begin{equation}
 \label{eq:lagrange_sol6_eq}
\alpha_1 =\frac{2\alpha_2 x_2 + \beta_2- \beta_1}{2x_1} 
\end{equation}

Equation \eqref{eq:lagrange_final_sol_eq} generalizes \eqref{eq:lagrange_sol6_eq} to calculate optimal $\hat{\alpha}_t$ values -- that is, values that cause customers to reproduce the pre-computed optimal $\hat{x}_t$ values -- using a selected seed load value $\hat{x}_{t_\text{seed}}$, associated $\beta_{t_\text{seed}}$ price, and $\alpha_{t_\text{seed}}$ value. We can select ${t_\text{seed}}$ and $\alpha_{t_\text{seed}}$ to minimize the cost to the customer. To do this, we select the time period $t$ with the highest $\beta_t$ and non-zero load \eqref{eq:lagrange_seed_eq}. For $\alpha_\text{seed}$, since the other $\hat{\alpha}_t$ values are calculated from $\alpha_\text{seed}$, we select a value that is as small as practical. In our analysis, we used $\alpha_\text{seed} = 0$. However, there may be issues with using $\alpha_\text{seed} = 0$ if the DSO incorrectly forecasts customer load. We discuss forecasting issues in Section \ref{sec:considerations}.   

\begin{equation}
 \label{eq:lagrange_final_sol_eq}
\hat{\alpha}_t =\frac{2\alpha_{t_\text{seed}} \hat{x}_{t_\text{seed}} + \beta_{t_\text{seed}}- \beta_t}{2\hat{x}_t} 
\end{equation}

\begin{equation}
 \label{eq:lagrange_seed_eq}
t_\text{seed} =\operatorname{argmax}_{t}(\beta_{\hat{x}_t > 0})
\end{equation}

Due to the structure of \eqref{eq:lagrange_final_sol_eq} and the non-negativity requirement of the quadratic optimization, we place two constraints on $\hat{\alpha}_t$ to ensure feasibility. In times $t$ when there is no load ($\hat{x}_t$ = 0), then $\hat{\alpha}_t$ would calculate to $\infty$. If the $\hat{\alpha}_t < 0$, then the optimization would become non-convex. In these situations, we replace $\hat{\alpha}_t$ with a finite, non-negative $\theta$ value set based on type of controllable load the customer owns using Eq.~\eqref{eq:alpha_protection}.

\begin{equation}
\label{eq:alpha_protection}
\alpha_t = \begin{cases}
\theta,
&\text{if }\hat{\alpha}_t = \pm\, \infty, \\
\theta, 
&\text{if }\hat{\alpha}_t < 0, \\
\hat{\alpha}_t 
&\text{else.}
\end{cases}
\end{equation}

If the controllable load the customer is optimizing is separately metered, then $\theta$ must be larger than the finite $\hat{\alpha}_t$ values calculated in \eqref{eq:lagrange_final_sol_eq} to ensure the customer follows the target load profile $\hat{x}_t$. In this case we set $\theta$ to an arbitrarily high value of 10 to ensure finite, non-negative $\hat{\alpha}_t$ values. (We examined performance for other high values of $\theta$ and did not observe meaningful differences in the solution.) 

\begin{equation}
 \label{eq:shared_meter_loads}
\hat{x}_t = x_t + \Tilde{x}_t 
\end{equation}

However, a high value for $\theta$ can cause a control issue for the DSO when customers have non-controllable loads ($x_t$) on the same meter as controllable loads ($\Tilde{x}_t$) \eqref{eq:shared_meter_loads}, and those controllable loads are bi-directional. If the DSO requires customers to inject energy ($\hat{x}_t < 0$) at times when $\beta_t \not= \max(\beta_t)$, then the DSO needs to ensure that injections ($\hat{x}_t < 0$) also occur at $t = \operatorname{argmax}_t(\beta_t)$. Otherwise, in those times $\alpha_t = \theta$ and customers will optimize to $\hat{x}_t = 0$.

In contrast, a special case occurs if the controllable load is unidirectional ($\Tilde{x}_t >0$) and on the same meter as non-controllable load \eqref{eq:shared_meter_loads}. In that situation, the DSO can reduces costs by ignoring $x_{t_\text{seed}}$ in $\hat{x}_{t_\text{seed}}$ when calculating $\hat{\alpha}_t$ in Eq.~\eqref{eq:lagrange_seed_eq}. Then setting $\theta = 0$ reduces $\hat{\alpha}_t$ for times $t$ where the optimal solution would not place controllable load ($\Tilde{x}_t = 0$) due to the $\beta_t$ price. This reduces the price for energy in times where only non-controllable load is consuming ($x_t >0$) to $\beta_t$. This does not work with bi-directional loads ($\Tilde{x}_t <0$), since they would not inject energy at the times prescribed in the target load profile ($\hat{x}_t < 0$).

By calculating $\hat{\alpha}_t$ from Eq.~\eqref{eq:lagrange_final_sol_eq} -- \eqref{eq:shared_meter_loads}, we produce the minimum feasible vector of $\alpha_t$ prices that can produce the vector of load profiles $x_t$ given the $\beta_t$ prices and the $\alpha_{t_\text{seed}}$.

\section{Inverse-Rank LRP}
\label{sec:IR-LRP}
\subsection{Concept}
While the Optimal-$\alpha$ LRP is the minimum cost formulation of Load Responsive Pricing, it requires a target load profile for customers to optimize towards. DSOs can still utilize LRP for congestion management with other formulations as well. The ``inverse-rank LRP" (IR-LRP) is a variation of the LRP tariff that manages congestion via a heuristic method to set $\alpha_t$. It works by increasing $\alpha_t$ inversely to the $\beta_t$ energy price. 
\vspace{-2.5mm}
\subsection{Method}
The method for calculating $\alpha_t$ values for IR-LRP assumes that price-optimizing controllable loads are most likely to cause congestion management issues at times when energy prices are the lowest in the day. %The $\alpha_t$ price is calculated for each pricing period $t$ in the set of $\beta_t\; \forall t \in T$ prices. Without loss of generality, we assume $t$ to be hourly over a $T = 24$ hour pricing day. 

First, the DSO defines an $n$-length $\boldsymbol{\tau}$ as vector of evenly spaced values on an interval $[\tau_{\text{min}},\tau_{\text{max}}]$. They then re-index the $\boldsymbol{\tau}$ vector such that the index of the smallest entry of $\boldsymbol{\tau}$ equals the index of the largest $\boldsymbol{\beta}$ entry, the index of the second smallest entry of $\boldsymbol{\tau}$ equals the index of the second largest entry of $\boldsymbol{\beta}$, and so on. 

For our case studies and in preliminary load control testing, we found an effective linear range for $\boldsymbol{\tau}$ is $[0.1,\tau_\text{max}]$ with $\tau_\text{max} \in [1,3]$ depending on how much load shifting is required by the DSO. The larger the value of $\tau_\text{max}$, the more load will be shifted away from the minimum $\beta_t$ period.

Then the $\alpha_t$ values are set by

\begin{equation}
 \label{eq:IR-LRP_eta_tau_eq}
\alpha_t = \tau_t \cdot \eta,
\end{equation}
%

% First, $\alpha_t$ is decomposed into a temporal component $\tau_t$ and a customer scaling factor $\eta$ \eqref{eq:IR-LRP_eta_tau_eq}. 
% Next, a $n$-length vector $\textbf{\mu}$ is defined by the DSO to create a linear range of multipliers. For our case studies and in preliminary load control testing, we found an effective linear range starts at $\mu_{1} = 0.1$ and ending at $\mu_{n} \in [1,3]$ depending on how much load shifting is required by the DSO. The larger the $\mu_{n}$, the more load will be shifted away from the minimum $\beta_t$ period.

% We calculate $\tau_t$ as the time sorted values of $\mu_i$. Where $\mu_i$ is sorted by inversely ranking $\beta_t$ periods, such that the $t$ time with the largest $\beta_t$ price is $\mu_1$ and the smallest $\beta_t$ is last $\mu_n$. See Table \ref{tab:Case_I_tau_values} of Case Study I to see an example of $\tau_t$ values.

\noindent where $\eta$ is a customer scaling factor corresponding with the size of their combined controllable and non-controllable load on the same meter. Each customer or customer class can have an $\eta$ set for their controllable load size at the time of adoption of the IR-LRP tariff. Without $\eta$ or with a single $\eta$ value, larger customers would see higher prices than smaller customers because they consume more non-controllable load. For example, a DSO can have an $\eta = 0.001$ for a customer class up to 100 kW and $\eta = 0.0001$ for a separate customer class up to 1000 kW. By using different $\eta$ values for different customer sizes, the DSO can control how much each customer is shifting their load and also ensuring costs rise at a rate commensurate with the customer's relative shift in load. 

\begin{figure}[htbp]
\centering
\includegraphics[width=3.25in, height = 2.25in]{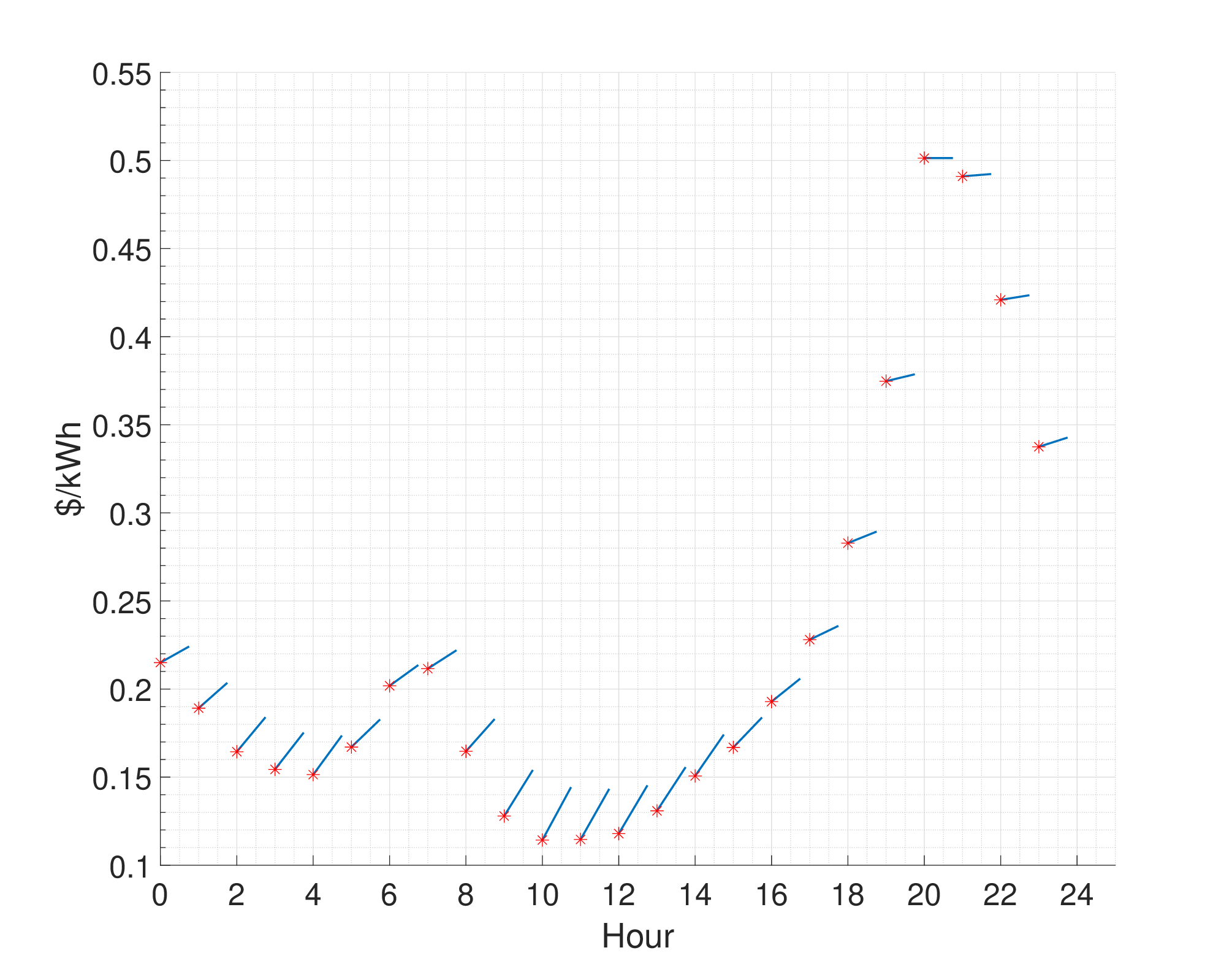}
\caption{Hypothetical hourly IR-LRP example with constant load. The intercept at each hour is the $\beta_t$ (red asterisk) while the $\alpha_t$ is the slope of the blue line. In this example, 10am is the lowest $\beta_t$ price and has the highest $\alpha_t$, $\alpha_t$ = 3$\eta$. While 8pm has the highest $\beta_t$ price and lowest $\alpha$, $\alpha$ = 0.}
\label{fig:IR-LRP_example}
\end{figure}

Once a DSO has assigned an $\eta$, $\tau_t$ can be calculated each time period and multiplied by $\eta$ to produce the set of $\alpha_t$ prices \eqref{eq:IR-LRP_eta_tau_eq}. Fig \ref{fig:IR-LRP_example} shows a hypothetical set of IR-LRP prices with a constant controllable load.

% These $\tau$ prices are then scaled by $\eta$ to correspond with the size of their combined controllable and non-controllable load on the same meter. Each customer or customer class can have an $\eta$ set for their load size at the time of adoption of the IR-LRP tariff. Without $\eta$, larger customers would see higher prices because they consume more load. For example, a DSO can have an $\eta = 0.001$ for a customer class up to 100 kW and $\eta = 0.0001$ for a separate customer class up to 1000 kW. 

% The algorithm sets $\alpha$ as the product of the inverse ranking of the $\beta$ prices ($\tau$), and the size of the load for the customer class ($\eta$). Conceptually, this heuristic is based on the idea OCLs are most likely to cause congestion management issues at the times with the lowest $\beta$ prices. Larger $\alpha$ coefficients will be necessary at those times to incentivize load curtailment at high consumption levels. To keep costs down for the customer, IR-LRP automatically bases the size of $\alpha$ inversely to the cost of $\beta$. There is an additional scaling factor as well to keep the cost of $\alpha$ appropriate for the size of the load. Customers can calculate their $\alpha$ values based on $\beta$ prices and $\eta$ scaling factors that are based on the customer's load size and type.

\section{Case Studies}
\label{sec:case-studies}
To compare the LRP approaches, we modeled two case studies. The first case study is a single building to examine the effectiveness of LRP to shape a customer's load profile. The second case study measures the effects of LRP for congestion management on a test system with high potential of network congestion. To show the diversity of applications for LRP, we use different $\beta_t$ price sources, load type (bi-directional vs. unidirectional), and connection (separately metered vs single meter) in each case study.

%CASE 1 - Single building with bi-directional loads (separately metered)
\vspace{-2.5mm}
\subsection{Case Study I}
This first case study shows how LRP can be used for price-signal control of a single customer. We model a customer's optimal load profile under a day-ahead energy pricing tariff, the IR-LRP, and the Optimal-$\alpha$ LRP. We performed the optimizations in MATLAB R2020a \cite{2020MATLAB} using YALMIP \cite{YALMIP} with Gurobi \cite{Gurobi} as the solver. All calculations and simulations were performed on an AMD Ryzen 7 3700X desktop PC with 32GB of RAM, running Windows 10.

In this model, we assume the customer has their controllable loads separately metered from their other building loads. Specifically, we assume the customer requires 60 kWh of energy at a maximum of 20 kW. The customer also has an energy storage system they want to use to sell 10 kWh of energy to the DSO at a maximum power of 10 kW. However, for this case study we assume the DSO wants to limit peak power for this customer to 15 kW. We also assume the DSO does not have any means of communication or control of the customer except the ability to send a set of $\alpha_t$ and $\beta_t$ prices.

Table \ref{tab:Case_I_beta_prices} lists the $\beta_t$ prices for this customer. We used a set of hourly volumetric energy prices designed by Lawrence Berkeley National Laboratory (LBNL) to provide full cost recovery to the utility \cite{LBNL_CalFuse_2022}. These prices are the hourly prices from the LBNL study for March 1st, 2019 in the San Diego Gas and Electric (SDG\&E) territory. We use these $\beta_t$ prices both for LRP tariffs and in the day-ahead energy pricing tariff to compare the customer's load profile in a hourly pricing program without LRP.

\begin{table}[htbp]
\caption{$\beta_t$ values for Case Study I}
\begin{center}
\begin{tabular}{c c c}
\begin{tabular}{c|c}
\hline
Hour & $\beta_t (\$/\text{kWh})$ \\
\hline
0 & 0.2198 \\
1 & 0.2074 \\
2 & 0.2044 \\
3 & 0.1945 \\
4 & 0.2081 \\
5 & 0.2632 \\
6 & 0.3349 \\
7 & 0.3226 \\
8  & 0.2318 \\
9  & 0.1773 \\
10 & 0.1479 \\
11 & 0.1397 \\
\hline
\end{tabular}
& \hspace{2mm} &
\begin{tabular}{c|c}
\hline
Hour & $\beta_t (\$/\text{kWh})$ \\
\hline
12 & 0.1455 \\
13 & 0.1630 \\
14 & 0.1711 \\
15 & 0.1839 \\
16 & 0.2739 \\
17 & 0.4124 \\
18 & 0.5185 \\
19 & 0.4680 \\
20 & 0.4213 \\
21 & 0.3841 \\
22 & 0.3393 \\
23 & 0.2833 \\
\hline
\end{tabular}
\end{tabular}
\end{center}
\label{tab:Case_I_beta_prices}
\end{table}

We model the customer behavior and energy billing at the 15-min timescale to match the measuring frequency of an SDG\&E business smart meter. However, since the prices in the case study are hourly, we report our results in hourly intervals.

\subsubsection{IR-LRP Parameters}
For this case, we assume the DSO has set $\eta = 0.001$ for the the IRP-LRP for this customer's customer class. Using these parameters for this customer and the $\beta_t$ prices, the IR-LRP first creates a set of $\tau_t$ values which then generate a set of $\alpha_t$ prices with the largest $\alpha_t$ at 11:00 and the smallest $\alpha_t$ at 18:00 (Table \ref{tab:Case_I_alpha_prices}).

\begin{table}[htbp]
    \caption{IR-LRP $\tau_t$ values in $[0.1, 1.5]$  and $\alpha_t$ $(10^{-4}\$/\text{kWh}^2)$ values for Case Study I}
    \begin{center}
    \begin{tabular}{c c c}
    \begin{tabular}{c|c|c}
    \hline
    Hour & $\tau_t$ & $\alpha_t$\\
    \hline
    0 & 0.83 &  8.30\\
    1 & 0.95 &  9.52\\
    2 & 1.01 & 10.13\\
    3 & 1.07 & 10.74\\
    4 & 0.89 &  8.91\\
    5 & 0.71 &  7.09\\
    6 & 0.47 &  4.65\\
    7 & 0.53 &  5.26\\
    8  & 0.77 &  7.70\\
    9  & 1.20 & 11.96\\
    10 & 1.38 & 13.78\\
    11 & 1.50 & 15.00\\
    \hline
    \end{tabular}
    & \hspace{-2mm} &
    \begin{tabular}{c|c|c}
    \hline
    Hour & $\tau_t$ & $\alpha_t$\\
    \hline
    12 & 1.44 & 14.39\\
    13 & 1.32 & 13.17\\
    14 & 1.26 & 12.57\\
    15 & 1.13 & 11.35\\
    16 & 0.65 & 6.48\\
    17 & 0.28 & 2.83\\
    18 & 0.10 & 1.00\\
    19 & 0.16 & 1.61\\
    20 & 0.22 & 2.22\\
    21 & 0.34 & 3.43\\
    22 & 0.40 & 4.04\\
    23 & 0.59 & 5.87\\
    \hline
    \end{tabular}
    \end{tabular}
    \end{center}
    \label{tab:Case_I_alpha_prices}
    \end{table}

\subsubsection{Optimal-\texorpdfstring{$\alpha$}{α} LRP Design}
In contrast to the IR-LRP where the DSO creates prices to curtail load at its peak, the Optimal-$\alpha$ LRP allows a DSO to incentivize a customer to follow a specific load profile. In this case, we assume the DSO knows the customer's energy needs and wants the customer to consume and discharge energy at the levels in Table \ref{tab:Case_I_optimal_alpha_values}. We make no assumptions on the approach the DSO took to calculate these values and the DSO does not communicate this target load profile to the customer. Instead, the DSO uses this target load profile to calculate $\alpha_t$ values for the customer to use in their optimization to recreate this load profile as their optimal solution.

\begin{table}[htbp]
    \caption{Optimal-$\alpha$ LRP target load profile (kWh) and $\alpha_t$ $(10^{-4}\$/\text{kWh}^2)$ values for Case Study I}
    \begin{center}
    \begin{tabular}{c c c}
    \begin{tabular}{c|c|c}
    \hline
    \makecell{Hour} & \makecell{kWh} & \makecell{ $\alpha_t$}\\
    \hline
    0  & 0.   & 10               \\
    1  & 0.   & 10               \\
    2  & 0.   & 10               \\
    3  & 0.   & 10               \\
    4  & 0.   & 10               \\
    5  & 0.   & 10               \\
    6  & 0.   & 10               \\
    7  & 0.   & 10               \\
    8  & 10.  & $1\times10^{-13}$\\
    9  & 2.   & 0.0136           \\
    10 & 12.  & 0.0035           \\
    11 & 15.  & 0.0031           \\
    \hline
    \end{tabular}
    & \hspace{-2mm} &
    \begin{tabular}{c|c|c}
    \hline
    \makecell{Hour} & \makecell{kWh} & \makecell{ $\alpha_t$}\\
    \hline
    12 & 13   & 0.0033 \\
    13 & 3.   & 0.0115 \\
    14 & 5.   & 0.0061 \\
    15 & 0.   & 10     \\
    16 & 0.   & 10     \\
    17 & 0.   & 10     \\
    18 & -10. & 0.0143 \\
    19 & 0.   & 10     \\
    20 & 0.   & 10     \\
    21 & 0.   & 10     \\
    22 & 0.   & 10     \\
    23 & 0.   & 10     \\
    \hline
    \end{tabular}
    \end{tabular}
    \end{center}
    \label{tab:Case_I_optimal_alpha_values}
    \end{table}

Using the Optimal-$\alpha$ formula, we calculate the optimal $\alpha_t$ prices for the customer (Table \ref{tab:Case_I_optimal_alpha_values}). Note that $\alpha_t = 10$ for hours when there is no load in the target load profile. As such, the customer's cost in those hours is \$0 when they optimize their load.

\subsubsection{Results}
\begin{figure}[htbp]
\centering
\includegraphics[width=3.4in]{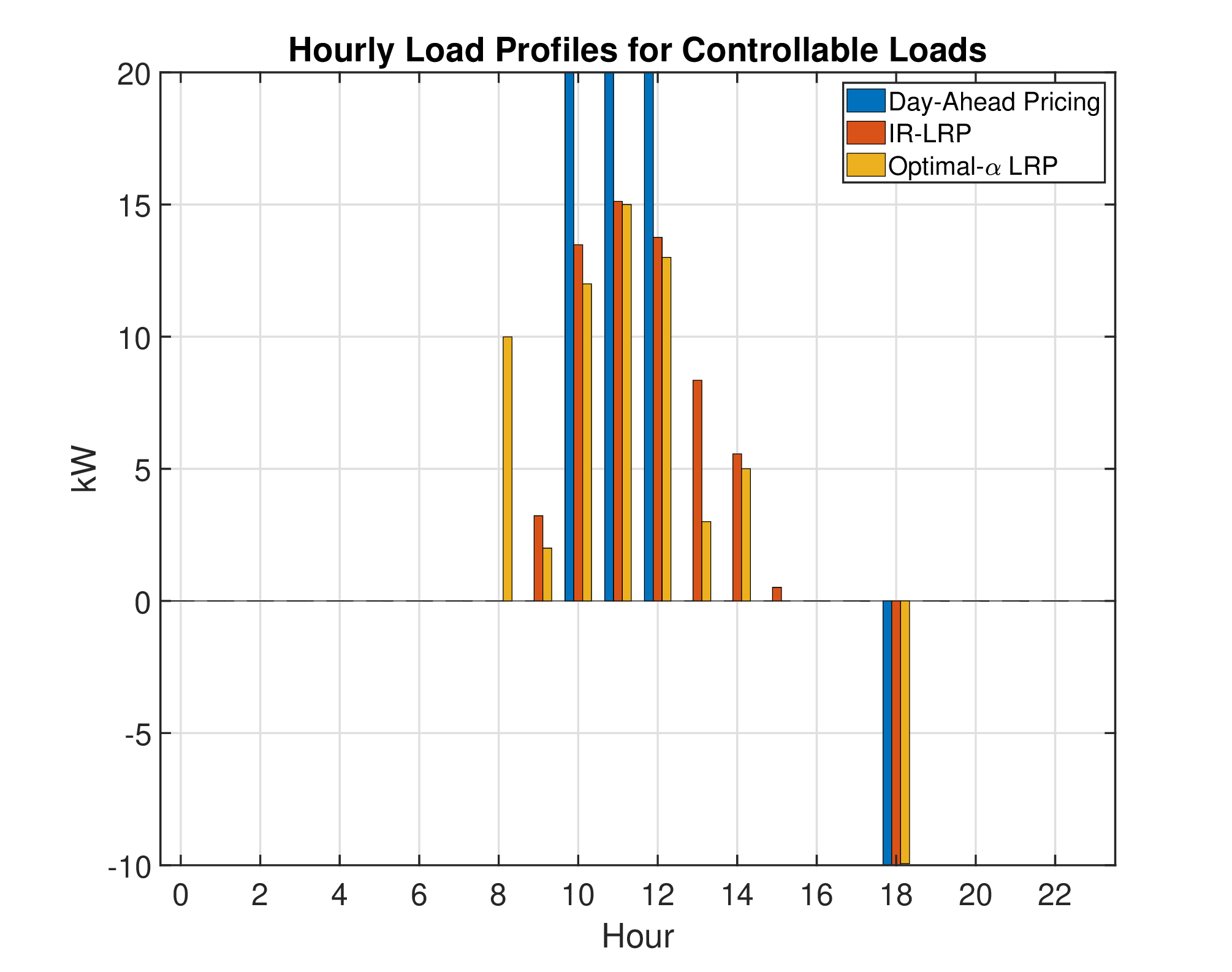}
\caption{Case Study I - Load profile for a customer following the day-ahead pricing tariff, IR-LRP, and Optimal-$\alpha$ LRP }
\label{fig:Case_Study_I_Load_Profile}
\end{figure}

Using the day-ahead pricing, IR-LRP, and Optimal-$\alpha$ LRP, we model the customer minimizing their costs in a 24-hour optimization. Figure \ref{fig:Case_Study_I_Load_Profile} shows the resulting load profiles for this customer following these three tariffs. Under day-ahead pricing, the customer maximizes their consumption up to their local load limit (20 kW) for the three lowest $\beta_t$ hours of the day. In contrast, under the IR-LRP tariff, the customer's load was spread across more hours. Finally, under the Optimal-$\alpha$ LRP tariff, the customer followed the target load profile of the DSO to within 0.06 kWh accuracy. This accuracy can be improved to 0.002 kWh when the optimization is performed in watt-hours instead of kilowatt-hours. These results show that LRP can effectively shift customer load.

\begin{figure}[htbp]
\centering
\includegraphics[width=3.4in]{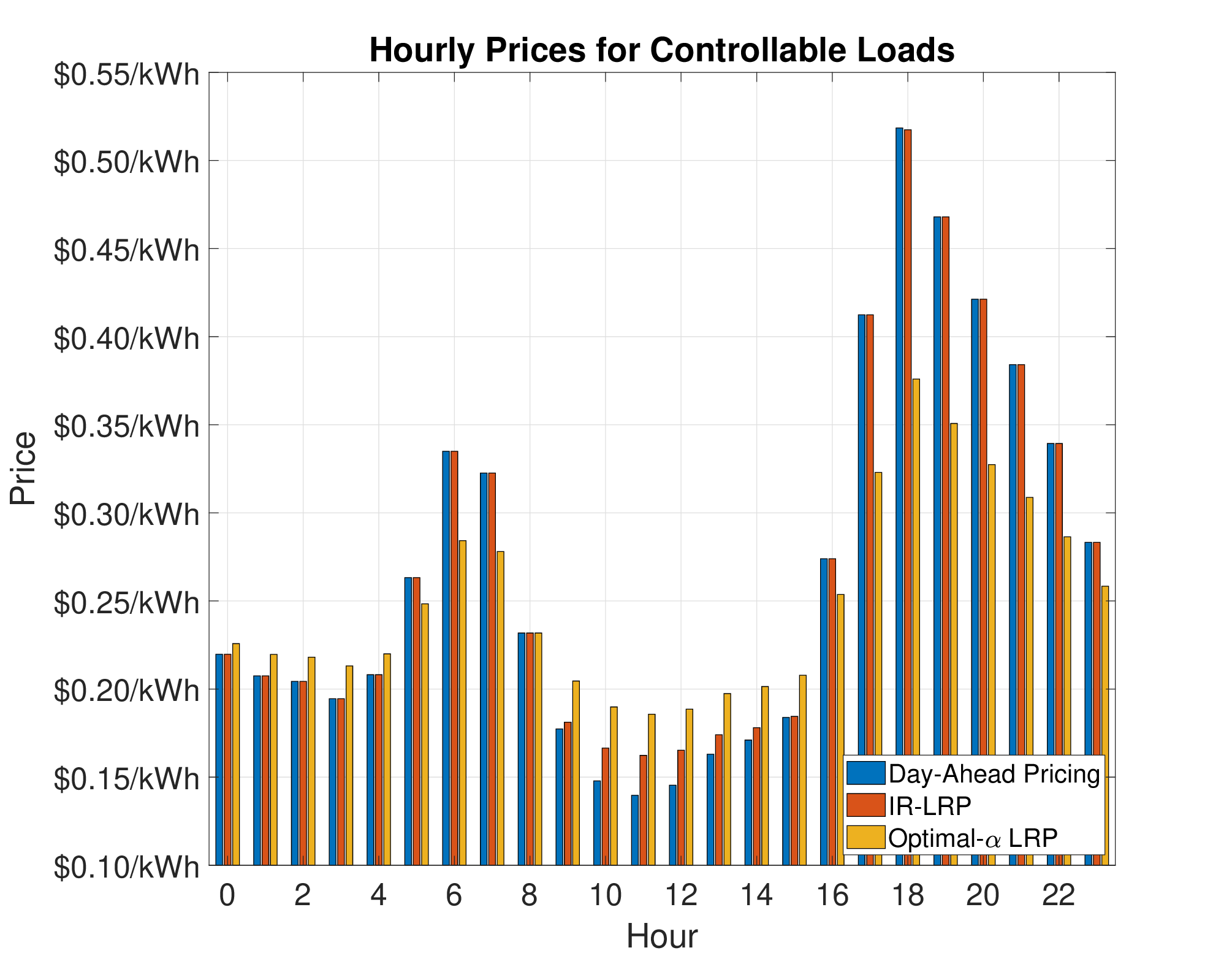}
\caption{Case Study I - Prices for a customer following the day-ahead pricing tariff, IR-LRP, and Optimal-$\alpha$ LRP }
\label{fig:Case_Study_I_Prices}
\end{figure}

Figure \ref{fig:Case_Study_I_Prices} shows the total marginal \$/kWh price at each hour for the three different tariffs. We calculate the IR-LRP and Optimal-$\alpha$ LRP \$/kWh at the resulting load levels from Figure \ref{fig:Case_Study_I_Load_Profile}. The IR-LRP had a maximum hourly price increase of \$0.0227/kWh over the day-ahead pricing tariff, and the Optimal-$\alpha$ LRP had a maximum price increase of \$0.0461/kWh. However, the price the customer would receive for the energy they sold back to the grid fell by \$0.1425/kWh. Over the course of a billing period, these changes would result in increased bills for the customer. We explore monthly billing effects of LRP in Case Study II and discuss several strategies a DSO could employ to mitigate these cost increases in Section \ref{sec:considerations}.

%CASE 2 - PG&E Feeder with EVs
\subsection{Case Study II}
To examine the congestion management capabilities of LRP, we modeled commercial customers with large fleets of electric vehicles (EVs) on a heavily loaded, three-phase unbalanced distribution feeder that is susceptible to undervoltage due to day-time charging of the EVs. In this model, we compared commercial customers optimizing their EV charging in a centralized optimization to a decentralized LRP approach where customers optimized only for their costs. We modeled the centralized optimization as a cost minimizing linear program under day-ahead hourly prices. 

We performed the optimizations in MATLAB R2020a \cite{2020MATLAB} using YALMIP \cite{YALMIP} with Gurobi \cite{Gurobi} as the solver and the distribution system powerflow simulations were modeled in Gridlab-D \cite{Chassin2008GridLAB-D}. All calculations and simulations were performed on an AMD Ryzen 7 3700X desktop PC with 32GB of RAM, running Windows 10.

\subsubsection{Model Design}
\begin{figure}[htbp]
\centering
\includegraphics[width=3in, height = 2.5in]{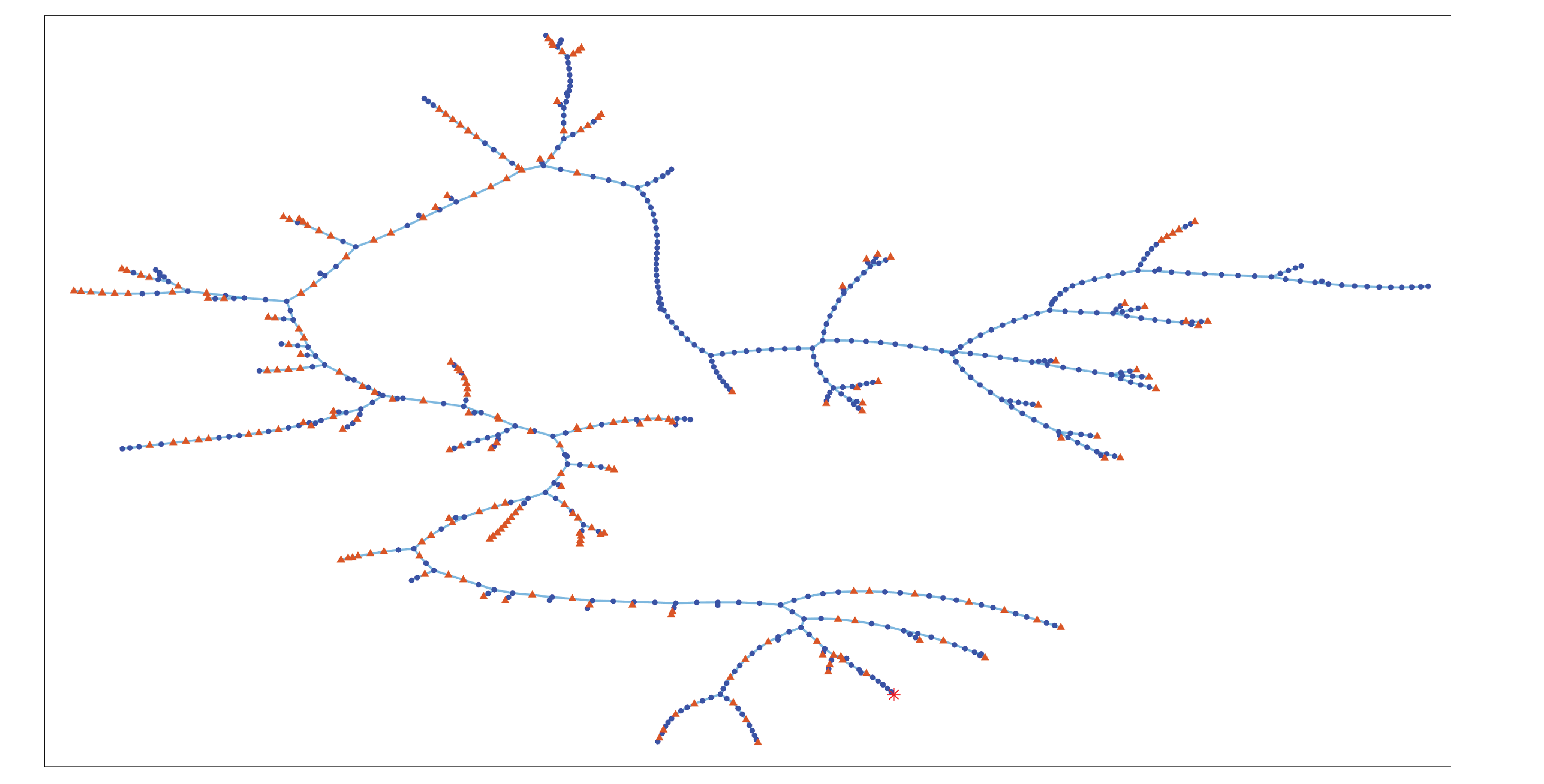}
\caption{Graph representation of the distribution feeder model with 678 nodes in a force-directed layout (identified as Feeder D0001 in \cite{PGE_CEC_2015}). The red star indicates the distribution substation and orange triangles are the 221 load locations identified in the original feeder model.}
\label{fig:PGE_Feeder}
\end{figure}

For our distribution feeder, we used a 17-mile, three-phase unbalanced distribution feeder created by PG\&E to model an existing urban distribution feeder in the inland area of Northern California \cite{PGE_CEC_2015}. The feeder has overhead and underground power lines supplying 2,894 residential customers, 270 commercial customers, and 91 industrial customers (Fig. \ref{fig:PGE_Feeder}). To provide load diversity, we modified the feeder model to incorporate 10 different building models from the Department of Energy’s Commercial Reference Building models and Building America House Simulation Protocols for the residential building models \cite{Deru2011U.S.}, \cite{Hendron2010Building}.

We optimized the EV charging of two types of three-phase customers, medium office building customers and warehouse customers, each with fleets of electric vehicles operating as flexible, price-optimizing loads. We modeled all 91 industrial customers on the feeder as warehouse buildings and five of the commercial customers as medium-sized office buildings. We modeled a total of 2,160 EVs on the distribution feeder, 10 EVs at each of the 91 warehouses and 250 EVs at each of the 5 office buildings, for a combined load of 43.2 MWh per day or 1.34 GWh per month. Before running our optimization, we verified that the feeder could support charging the total number of EVs we modeled without voltage constraint violations if the vehicles did not all charge in the same time period.

For prices, we used hourly prices from the Avoided Cost Calculator developed by the energy consulting firm E3 for the California Public Utilities Commission to determine the benefit of distributed energy resources \cite{ACC_2022_Resolution}. The ACC calculates the avoided cost of energy for every hour of the year by modeling the combined costs of wholesale energy, system capacity, and environmental damages from greenhouse gases. Specifically, we use the 2022 edition of ACC electricity prices for the PG\&E climate area 12 in the year 2023 \cite{ACC}. We used the ACC prices as both a measure of the total social cost and as a day-ahead hourly price for the office building and warehouse customers.

\subsubsection{Centralized Optimization}
To simulate the centralized optimization, we developed a linear programming model to minimize the daily cost of charging a fleet or collection of electric vehicles at customer locations \eqref{eq:EV_optimization}.  Where $\pi_t$ is the ACC price at time $t$, $\Tilde{x}_{i,b,t}$ is the sum of all EV loads and $x_{i,b,t}$ is the sum of all non-controllable building loads at building $b$ in set of $B_i$ buildings at node $i$ the set of nodes $I$ at time $t$. The local loading limit prevent coincidental load from exceeding the circuit breaker limit of each customer. Real power from a single EV $l$ in the set of all EVs $L_b$ at a building is represented by $w_{l,i,b,t}$ and is constrained by the EV operating constraints, which are the maximum charging rates of each EV (7.2kW) with only unidirectional flow from the grid to charge the EV. We modeled each EV requiring 20kWh/day to model an extreme case for EV charging. We also required each EV to be fully charged at the end of each day.

\begin{align}
\label{eq:EV_optimization}
\min_{\Tilde{x}}{\text{Cost}} = \sum_{t=1}^{24\text{ hours}}\sum_{i=1}^{I}\sum_{b=1}^{B_i}&\pi_t\cdot\left(\Tilde{x}_{i,b,t} + x_{i,b,t}\right) \\ 
s.t.\quad\left(\Tilde{x}_{i,b,t} + x_{i,b,t}\right) &\leq \text{Local loading limit} \nonumber \\
\quad \Tilde{x}_{i,b,t} &=\sum_{l=1}^{L_b}{w_{l,i,b,t}}  \nonumber \\
\quad w_{l,i,b,t} &\leq \text{EV operating constraints} \nonumber \\
\quad \sum_{t=1}^{24\text{hrs}}{\Tilde{x}_{i,b,t}} &= \text{charging energy required for } 
\nonumber\\[-4mm]
& \quad \;\text{EVs at the building} \nonumber
\end{align}

The linear program minimizes the total cost to the customer for their building loads and the charging of all the electric vehicles at the customer’s site. In our analysis, we only optimized EV charging; the building loads were treated as exogenous to the optimization and only included for their impact on the voltage constraint and costs. We repeated this daily optimization for every day in the month of July.

We performed the centralized optimization first without a voltage constraint and then with a linearized voltage magnitude constraint \eqref{eq:LinDist3Flow} and \eqref{eq:LinDist3Flowconstraint} based on the LinDistFlow equations developed by Baran and Wu \cite{Baran_Wu_1989} and extended into three-phase versions in \cite{Gan_Low_2014} and \cite{Sankur_Dobbe_Stewart_Callaway_Arnold_2016}. This LinDisFlow (LDF) constraint ensured voltage magnitude at all nodes in the centralized optimization was kept at or above 0.95pu as required by ANSI C84.1 \cite{ANSI_C84_1_2016}.

\begin{align}
\label{eq:LinDist3Flow}
|V_{\theta_{k},t}|^2 \approx |V_{\theta_{i},t}|^2-2(&R_{\theta_{k}}(\boldsymbol{P}_t+\boldsymbol{\Tilde{P}}_t) +X_{\theta_{k}}(\boldsymbol{Q}_t+\boldsymbol{\Tilde{Q}}_t)) \nonumber \\
& \forall i,k \in I 
\end{align}

\begin{equation}
\label{eq:LinDist3Flowconstraint}
 |V_{\theta_{k},t}|^2 \geq  |V_{min}|^2 \quad \forall k \in I, \forall t \in \text{24 hours,}
\end{equation}
where $|V_{\theta_{i},t}|$, $|V_{\theta_{k},t}|$ are voltages at each phase $\theta$ at node $i$ and $k$ respectively at time $t$, and $i,k$ denote nodes in set $I$ of all nodes for the feeder. The line resistance and reactance for each phase $\theta$ from $i$ to $k$ are $R_{\theta_{k}}$, $X_{\theta_{k}}$ respectively. $\boldsymbol{P}_t$ and $\boldsymbol{Q}_t$ are the vectors of uncontrollable building load real and reactive power at each phase $\theta$ of the load nodes respectively, while $\boldsymbol{\Tilde{P}}_t$ and $\boldsymbol{\Tilde{Q}}_t$ are the vectors of controllable EV load real and reactive power at time $t$. Building loads were assumed to have a power factor of 0.9 and EVs were assumed to have unity power factor. All loads were modeled as constant power loads and all loads at three-phase connected buildings were modeled as balanced three-phase loads across each phase $\theta$ \eqref{eq:P_to_x_mapping}. $V_\text{min}$ is the ANSI C84.1 voltage minimum value for delivered energy (0.95pu).

\begin{align}
\label{eq:P_to_x_mapping}
 P_{{\theta_i},t} = \frac{1}{3}\sum_{b=1}^{B}(x_{i,b,t}), &\quad Q_{{\theta_i},t} = P_{{\theta_i},t} \times \tan(\cos^{-1}(0.9)) \\
 \Tilde{P}_{{\theta_i},t} = \frac{1}{3}\sum_{b=1}^{B}(\Tilde{x}_{i,b,t}), &\quad \Tilde{Q}_{{\theta_i},t} = 0 \nonumber
\end{align}

However, customers in a central optimization with a voltage constraint would experience different costs based on their location in the distribution feeder. Customers in more congested areas might need to shift more load into higher cost hours than customers in less congested areas. While we leave equity considerations of this to policymakers, in our optimization all EV load of the same building model type are optimized to have the same load profile to allow for a simple comparison of customer bills across the distribution feeder.

\subsubsection{LRP Designs}
For the LRP tariffs, we had each customer optimize against their $\alpha_t$ and $\beta_t$ prices. The $\beta_t$ prices were the same ACC prices used in the centralized optimization. While the $\alpha_t$ was calculated using the  LRP formulas.

Table \ref{tab:caseII_IR_LRP_Design} lists the IR-LRP values used in this case study. The $\eta$ scaling factors for medium office building and warehouse building models are equivalent to \$0.0001/100kWh and \$0.0001/10kWh. These $\eta$ values were based on the order of magnitude of the non-controllable load at each building model.

\begin{table}[htbp]
\caption{IR-LRP $\tau_t$ and $\eta$ values for Case Study II\label{tab:caseII_IR_LRP_Design}}
\centering
\begin{tabular}{cccc}
\hline
\makecell{$\tau_t$ range} & \makecell{Office\\Building $\eta$}	& \makecell{Warehouse $\eta$} \\
\hline
 & & \\[-6pt]
 [0.1, 3] & $1\times 10^{-6}$ & $1\times 10^{-5}$ \\[3pt]
\hline
\end{tabular}
\vspace{-2mm}
\end{table}

For the Optimal-$\alpha$ LRP, we used the EV load profiles from the centralized optimization with the LDF constraint as target load profiles. The maximum absolute difference between the target load profiles and those generated by the optimal-$\alpha$ LRP was $3.57\times10^{-4}$ kWh. The resulting voltage profiles when modeled in Gridlab-D had a maximum absolute difference of $8.97\times10^{-6}$ V. Since the EV loads were modeled as unidirectional loads and on the same meter as the building loads, we used the maximum $\beta_t$ hour with EV load from our target load profiles when calculating the $\alpha_{\text{seed}}$ using \eqref{eq:lagrange_seed_eq}. This allowed $\alpha_t = 0$ for all hours $t$ where $\Tilde{x}_{i,b,t} = 0$.

\subsubsection{Results}
Table \ref{tab:caseII_voltages} lists the number of voltage violations over the month and maximum violation for each of the optimization scenarios. Table \ref{tab:caseII_customer_costs} shows the volumetric cost to an individual customer for following each of the different tariffs and social cost of all customers on the feeder optimizing against each tariff (measured in ACC costs). 

As expected, the ACC optimization without a voltage constraint caused severe voltage violations multiple times throughout the month. While the 9 days of voltage violations might be mitigated with CPP programs, since most CPP programs limit CPP events to less than 20 per year, it may not be possible to control voltage over the entire year with CPP programs only. However, the costs of the ACC optimization provides a cost basis to compare the other tariff designs.

The centralized optimization with the LDF constraint and both LRP tariffs kept voltage above 0.95pu. The centralized optimization with LDF constraint tariff resulted in the lowest cost for a customer. However, from the social cost perspective, the IR-LRP cost less across the entire distribution feeder. We believe this was due to the linear nature of the LDF constraint limiting the acceptable voltage range. If the voltage constraint in the centralized optimization could exactly model voltage in the distribution feeder, then the centralized optimization with this voltage constraint would have a lower social cost.

\begin{table}[htbp]
\caption{The minimum voltage measured across all customer locations in July, and the number of days in the month the voltage is below the ANSI C84.1 lower limit for voltage in Case Study II \label{tab:caseII_voltages}}
\centering
\begin{tabular}{ccc}
\hline
\makecell{Tariff} & \makecell{Minimum Voltage\\(Per Unit)}	& \makecell{Number of days\\voltage $<$ 0.95\;pu}\\
\hline
 & & \\[-6pt]
 ACC 	                   & 0.9265 & 9\\[3pt]
 ACC w/ LDF Constraint      & 0.9524 & 0\\[3pt]
 Optimal-$\alpha$ LRP       & 0.9524 & 0\\[3pt]
 IR-LRP                     & 0.9563 & 0\\[3pt]
\hline
\end{tabular}
\vspace{-5mm}
\end{table}
% \begin{table}[h]
% \caption{The monthly volumetric costs for an office building customer or warehouse customer under each tariff \label{tab:caseII_customer_costs}}
% \centering
% \begin{tabular}{ccc}
% \hline
% \makecell{Tariff} & \makecell{Office\\Building}	& \makecell{Warehouse}\\
% \hline
%  & & \\[-6pt]
%  ACC 	                   & \$10,940.59 & \$1,638.69\\[3pt]
%  ACC w/ LDF Constraint      & \$11,068.12 & \$1,649.60\\[3pt]
%  Optimal-$\alpha$ LRP (LDF) & \$11,459.05 & \$1,712.87\\[3pt]
%  IR-LRP                     & \$11,538.82 & \$1,681.13\\[3pt]
%  Optimal-$\alpha$ IR-LRP    & \$11,413.05 & \$1,666.97\\[3pt]

% \hline
% \end{tabular}
% \end{table}

\begin{table}[htbp]
\caption{The monthly volumetric costs when optimizing under each tariff and the social cost (in ACC dollars) of all customers on the distribution feeder in Case Study II \label{tab:caseII_customer_costs}}
\centering
\begin{tabular}{cccc}
\hline
\makecell{Tariff} & \makecell{Office\\Building}	& \makecell{Warehouse}  & \makecell{Social\\Cost}\\
\hline
 & & &\\[-6pt]
 ACC 	                   & \$10,940.59 & \$1,638.69 & \$203,824.11\\[3pt]
 ACC w/ LDF Constraint      & \$11,068.12 & \$1,649.60 & \$205,453.89\\[3pt]
 Optimal-$\alpha$ LRP       & \$11,439.58 & \$1,712.87 & \$205,453.89\\[3pt]
 IR-LRP                     & \$11,538.82 & \$1,681.13 & \$204,455.22\\[3pt]
\hline
\end{tabular}
\vspace{-2mm}
\end{table}

Table \ref{tab:caseII_percent_diff} lists the percent difference between these costs and the costs of optimizing against the ACC tariff without constraints. While the LRP tariffs did increase costs, as expected from a price-signal control approach, the cost increase were less than 10\% when compared to the centralized optimization that resulted in voltage violations. A question for rate-makers would be if the value of a decentralized congestion management would more than the  precent increase in cost for customers. In the following section we explore this issue and provide recommendations for reducing costs to the customer.

\begin{table}[htbp]
\caption{The percent difference between the ACC tariff and other tariff monthly volumetric costs in Case Study II \label{tab:caseII_percent_diff}}
\centering
\begin{tabular}{cccc}
\hline
\makecell{Tariff} & \makecell{Office\\Building}	& \makecell{Warehouse} & \makecell{Social\\Cost}\\
\hline
 & & & \\[-6pt]
 ACC                        & ---    &  ---   & ---    \\[3pt]
 ACC w/ LDF Constraint      & 1.17\% & 0.67\% & 0.80\%  \\[3pt]
 Optimal-$\alpha$ LRP       & 4.56\% & 4.53\% & 0.80\%  \\[3pt]
 IR-LRP                     & 5.47\% & 2.59\% & 0.31\% \\[3pt]

\hline
\end{tabular}
\end{table}
\vspace{-5mm}

\section{Implementation Considerations}
\label{sec:considerations}
While we have shown the LRP approach is an effective price-signal for congestion management, there are several potential issues that a DSO would need to consider before adopting this tariff. In this section, we discuss the real-world considerations of the LRP tariff design. We start with general considerations of an LRP tariff and then discussion considerations of both variations of the LRP tariff provided in this paper. 

First, the LRP tariff is complex compared to standard two-part tariffs with constant marginal costs. The LRP tariff is designed to influence the behavior of automated devices and works best when the customer is able to employ a quadratic optimization solver. This tariff should not be used as the default tariff for customers unless they have a sophisticated level of automation and control over their loads.

Second, this tariff assumes the $\beta_t$ prices are accurate price signals and that $\beta_t$ prices would be sufficient signals if network congestion was not an issue. If customers are not consuming in the preferred time period of the DSO, the DSO should first adjust $\beta_t$ prices to incentivize consumption during those hours. Only after the $\beta_t$ price is set for each time period should the $\alpha_t$ prices be set to create a soft cap incentive on consumption. The LRP tariff can compensate if $\beta_t$ prices are incentivizing for times that are problematic for the DSO but this comes at the cost of high $\alpha_t$ prices for customers, leading to the next consideration.

Third, LRP can have important cost impacts. We propose that this tariff should not capture more revenue than a real-time tariff and a well designed LRP tariff can minimize the cost impacts on consumers. However, a poorly designed tariff can cause significant cost increases for customers since the costs grow quadratically with load. The DSO or regulator will need to carefully set $\boldsymbol{\alpha}$ and $\boldsymbol{\beta}$ for each customer or customer class to prevent excessive cost burdens. In addition, revenue balancing mechanisms may need to be deployed. Baseline load profiles like those used in the real-time pricing scheme at Georgia Power, subscription plans, or annual credits to return excess revenue to customers could all be explored as ways to keep any sales generated by the $\boldsymbol{\alpha}$ coefficient revenue neutral.

Fourth, we have shown two different ways to calculate $\alpha_t$ prices for an LRP tariff. There are potentially other methods as well. Prices could be set with a granularity as low as the specific customer at a single location on a distribution feeder or as broad as a customer class in a utility service territory. The one key consideration for calculating $\alpha_t$ values is that they should always be non-negative for unidirectional loads or positive for bi-directional loads to preserve the quadratic formulation of the customer's optimization problem. 

Finally, the LRP tariff only provides a price for real power. In the future, DSOs may want to incentivize customers to provide reactive power. The LRP tariff could be used to incentivize reactive power but this would require putting a price reactive power and is beyond the scope of this paper.

\subsection{Optimal-\texorpdfstring{$\alpha$}{α} LRP Considerations}
The optimal-$\alpha$ LRP calculates the minimum $\boldsymbol{\alpha}$ required at each time period for customers to shift load to follow a target profile determined through the DSO's optimization. However, there are several issues that can arise with the optimal-$\alpha$ LRP. First, as seen in Case Study II, optimal-$\alpha$ LRP does not guarantee the the total costs for the customer will be the minimum cost possible. If the target load profile is not optimal, the results of Optimal-$\alpha$ LRP will not be either.

Second, the optimal-$\alpha$ LRP depends on an accurate forecast to calculate $\alpha_t$ values. The structure of LRP ensures that customers will still optimize their load if the forecast is wrong however different issues arise if the DSO overestimates vs. underestimates their forecast. 

If the DSO overestimates customer load in their forecast, customers will follow the general load shape the DSO intended with $\boldsymbol{\alpha}$ at a lower energy consumption at each time period. However, if a DSO underestimates load in their forecast, congestion issues can quickly arise. One solution to this issue is to always overestimate the expected load in the forecast. From a congestion management perspective, this is not a problem. Overestimating load will only cause congestion if the overestimated load level would be enough to cause congestion.

In cases where overestimating is not an option or if customer load exceeds the overestimated forecast, the customer a DSO underestimated will follow the load shape the DSO intended unless in any time period $\alpha_t = 0$. In those periods customers only optimize against $\beta_t$ and will consume energy without a linear price on quantity. As such, the majority of overestimated load will occur when $\alpha_t = 0$. One way to mitigate this issue is to assign any $\hat{\alpha}_{t} = 0$ a nonzero value after the calculating the $\hat{\alpha}_{t}$ prices. For example, if a DSO uses $\alpha_{t_\text{seed}} = 0$, then after calculating the $\hat{\alpha}_{t}$ values, the DSO could set $\alpha_{t_\text{seed}} = \min(\hat{\alpha}_{t})$. Then overestimated load will not cluster at $\alpha_{t_\text{seed}}$. 

This can lead to suboptimal $\alpha_t$ values if the seed is poorly selected (e.g. too large or at a time with no load). A way to mitigate this issue would be to formulate the $\boldsymbol{\alpha}$ calculation as a bi-level optimization problem with minimizing $\boldsymbol{\alpha}$ as the upper-level optimization task and minimizing costs to the customer as the lower-level task. However, this significantly increases the computational complexity of the algorithm. The seeding method is faster and sufficient for the size of the problem space for $\boldsymbol{\alpha}$.

Second, the optimal-$\alpha$ LRP depends on a seed $\alpha$ at a specific time to calculate $\alpha$ values. This can lead to suboptimal $\alpha$ values if the seed is poorly selected (e.g. too large or at a time with no load). A way to mitigate this issue would be to formulate the $\alpha$ calculation as a bi-level optimization problem with minimizing $\alpha$ as the upper-level optimization task and minimizing costs to the customer as the lower-level task. However, this significantly increases the computational complexity of the algorithm. The seeding method is faster and sufficient for the size of the problem space for $\alpha$.

Third, the Optimal-$\alpha$ LRP calculates $\alpha_t$ prices assuming that customers can accurately follow their target load profiles provided by the DSO. If customers have binding constraints in their optimization preventing this, the DSO's optimization should be updated to endogenize these constraints and recalculate the target profiles. 

Finally, since the DSO calculates prices based on the optimal load profile for each customer, the DSO could measure a customer's deviation from this ideal profile. This deviation could be the basis for performance metrics that compensate the customer for participating in the LRP tariff. 

\subsection{IR-LRP Considerations}
The IR-LRP is an effective heuristic pricing tariff for near communication-free congestion management that does not require any additional sensing or control equipment to deploy. However, it has several drawbacks and considerations that are important to highlight. First, the algorithm is a heuristic that assumes the $\beta_t$ prices will cause congestion management issues that need to be mitigated. The IR-LRP does not have knowledge of the status of the grid or customer loads. Instead, the IR-LRP assumes that the worst congestion will occur at the lowest $\beta_t$ prices.

Second, the IR-LRP algorithm is effective for over-consumption at low $\beta_t$ price times but is not effective at controlling bi-directional loads at high cost hours. The $\alpha_t$ calculation does provide incentives to curtail solar or discharging of energy storage back onto the grid during expensive $\beta_t$ price hours, and this could lead to over-voltage situations. In those circumstances, changing the $\beta_t$ price is the best strategy to prevent over-voltage. 

Finally, with more advanced sensing and grid awareness, a DSO could deploy IR-LRP with dynamic $\boldsymbol{\tau}$ and $\eta$ values to improve load control and reduce costs. However, if a DSO has this ability, other price setting algorithms may be more effective to use in the first place (such as the optimal-$\alpha$ LRP).

\section{Conclusion}
\label{sec:conclusion}
In this paper we describe a new linear pricing method, Load Responsive Pricing (LRP), for congestion management of price-signal controlled loads. We detail two different methods for calculating prices with LRP and show in case studies how customers would follow these price signals. We also show how this control method can alleviate network congestion. This work provides an alternative to previous price signal control research by leveraging other load optimization schemes beyond DLMP and provides a heuristic for deployment into the real-world without bi-directional communication or control. 

While we proposed the LRP tariff for day-ahead prices, an LRP approach could be used in other applications. One application could be as a control signal for non-monetary systems, such as an internal price-signal for a microgrid where decentralized optimization would reduce the computational burden of the central energy management system. Another application could be a sub-hourly $\alpha$ prices with hourly $\beta$ prices. This would allow a DSO to create a sub-hourly ramp to reduce hourly discontinuities in aggregate load that may occur under hourly pricing. We intend to explore both areas in future research.

As DSOs and regulators explore real-time pricing options and customer loads become more flexible with smarter optimization strategies, congestion management will become a more important issue for grid operations. Traditional volumetric energy pricing cannot alleviate congestion without real-time sensing, optimization, and communication of prices. LRP provides an additional control vector that traditional volumetric energy pricing does not have with a negligible increase communications and a potential revenue-neutral increase in customer costs. 

% {\appendix[Case Study Values]
% \label{sec:appendix}
% \input{Content/8_appendix.tex}
% }

\printbibliography
%

% \newpage


% \begin{IEEEbiography}[{\includegraphics[width=1in,height=1.25in,clip,keepaspectratio]{Content/Figures/Phanivong_small.eps}}]{Phillippe K. Phanivong}
%  (M’11) received the B.S. degree in electrical engineering from the University of Washington, Seattle, WA, USA, in 2015. He is currently pursuing the Ph.D. degree in energy and resources with the University of California, Berkeley, Berkeley, CA, USA.
 
% He served six years in the U.S. Navy as a Nuclear Power Plant Operator and Electrician. His current research interests include urban grid resiliency, integration of DERs, energy markets, and distribution system modeling techniques.
% \end{IEEEbiography}

% \begin{IEEEbiography}[{\includegraphics[width=1in,height=1.25in,clip,keepaspectratio]{Content/Figures/Callaway_small.eps}}]{Duncan S. Callaway}
%  (M’08) received the B.S. degree in mechanical engineering from the University of Rochester, Rochester, NY, USA, in 1995, and the Ph.D. degree in theoretical and applied mechanics from Cornell University, Ithaca, NY, USA.
 
% He is currently an Associate Professor of energy and resources with University of California, Berkeley, Berkeley, CA, USA. His current research interests include control strategies for demand response, electric vehicles and energy storage; distribution network management; and pathways to electrify and decarbonize the world’s low income regions.
% \end{IEEEbiography}


% \vfill

\end{document}